\documentclass[aps,prb,floatfix,amsfonts,amssymb,amsmath,twocolumn,showpacs,preprintnumbers,nofootinbib]{revtex4}
\usepackage{graphicx}
\usepackage{amsmath}
\def\pfhv{\hat{\mbox{\boldmath$p$}}_{\!\!{\scriptscriptstyle F}}}

\def\vR{{\mbox{\boldmath$R$}}}

\def\vd{{\mbox{\boldmath$d$}}}

\def\vh{{\mbox{\boldmath$h$}}}

\def\vq{{\mbox{\boldmath$q$}}}
\def\vp{{\mbox{\boldmath$p$}}}
\def\vpF{{\mbox{\boldmath$p$}}_{\!\!{\scriptscriptstyle F}}}

\def\vs{{\mbox{\boldmath$s$}}}

\def\vvF{{\mbox{\boldmath$v$}}_{\!\!{\scriptscriptstyle F}}}

\def\vR{{\mbox{\boldmath$R$}}}

\def\vgamma{{\mbox{\boldmath$\gamma$}}}
\def\vsigma{{\mbox{\boldmath$\sigma$}}}
\def\vDelta{{\mbox{\boldmath$\Delta$}}}
\def\tDelta{{\tilde\Delta}}

\begin{document}

\title{Structure of the core of magnetic vortices in d-wave superconductors
   with a subdominant triplet pairing mechanism}

\author{Mikael~Fogelstr\"om}Ê
\affiliation{Department of Microtechnology and Nanoscience, S-41296 G\"oteborg, Sweden}
\begin{abstract}
The quasiparticle states found in the vortex core of a high-T$_{\rm{c}}$ cuprate superconductor may be probed by scanning tunneling spectroscopy. Results of such experiments have revealed typical spectra that are quite different from what is seen in conventional low-Tc superconductors. In particular the Caroli-deGennes-Matricon state at $E\sim 0$ 
in the core center is not seen. Instead, in a high-T$_{\rm{c}}$ vortex core, quasiparticle states are found at energies that are at a sizable fraction of the gap energy. 
One explanation for this could be that a finite amplitude of a competing order parameter stabilizes in the vortex-core center. Here
I will explore the possibility 
of nucleating a vortex-core state that locally breaks inversion symmetry. The vortex-core order parameter is of mixed parity, 
aÊ$\,[d +i p]$-wave, 
and the quasiparticle spectra in the core center lacks the $E=0$ states.
\end{abstract}
\maketitle

\section{Introduction}
In conventional, nominally clean, type-II superconductors the quasiparticle spectrum in a vortex core was described by Caroli, deGennes, and Matricon. \cite{caroli1964} They found that quasiparticle states are localized in the vortex core, and that these same states carry the currents that screen the
magnetic flux line from penetrating the interior of the superconductor.
A direct measurement of vortex-core states was done by scanning tunneling spectroscopy (STS) \cite{hess1989} and the measured data could be {\em quantitatively} explained by theory. \cite{gygi1991} Performing 
STS measurements on vortices in high-T$_c$ cuprates, on YBCO \cite{maggioaprile1995} or on 
BSCCO \cite{pan2000,hoffman2002,levy2005}
(see also references in Ref. [\onlinecite{fischer2007}]) revealed a very different generic picture; the vortex core in a high-T$_c$ superconductor
does not harbor pronounced core states. By theory \cite{wang1995,schopohl1995,ichioka1996a,ichioka1996b} these core states should be present also for an order parameter of d-wave  symmetry relevant for high-T$_c$ superconductors and thus readily be seen in STS. Instead, 
non-dispersing vortex states are seen at energies corresponding  to $\sim30$\% of bulk value of the superconducting energy gap, $\Delta_0$.

Theoretical suggestions to explain the vortex-core spectra seen in high-T$_c$ cuprates include
possible subdominant singlet-paring amplitudes,\cite{ichioka1996a,ichioka1996b,franz1998} anti-ferromagnetic order stabilizing in the vortex core, 
\cite{arovas1997,andersen2000} the normal state pseudogap phase made visible in the core.\cite{berthod2001}, or competition
between d-density wave and d-wave superconducting order.\cite{maska2003}  
In this paper I will explore the possibility of nucleating a vortex-core state of mixed parity and
show that this state may have a finite amplitude of a p-wave order parameter in the vortex center.  The possibility of a mixed singlet-triplet pairing state 
in a vortex core has been suggested to occur in an s-wave superconductor as a result of spin-orbit coupling. \cite{salomaavolovik1987} 
In the context of a d-wave superconductors both spin-orbit coupling and the presence of a Zeeman coupling has been considered, finding a singlet-triplet mixing in the vortex phase in the Ginzburg-Landau region ($0\ll T\lesssim T_{\rm c}$).\cite{lebed2006,kabanov2007,dutta2008} Here, I will use the microscopic
quasiclassical theory to show, given an attractive pairing interaction in a p-wave channel and a weak intrinsic Zeeman coupling to the magnetic field to break the spin-singlet symmetry of the parent d-wave supeconductor, that a sizable p-wave order parameter may stabilize locally in the d-wave vortex core region. 
Using this {\em self-consistently} determined order-parameter field I then compute
the spatially resolved local density of states in the vortex. It turns out that the quasiparticle spectra seen in 
the STS on the high-T$_c$ cuprates \cite{maggioaprile1995,pan2000,hoffman2002,levy2005,fischer2007} can to large extent
be reproduced theoretically as a direct consequence of this triplet superconducting core order.   

\begin{figure*}[t]
\includegraphics[width=1.85\columnwidth,angle=0]{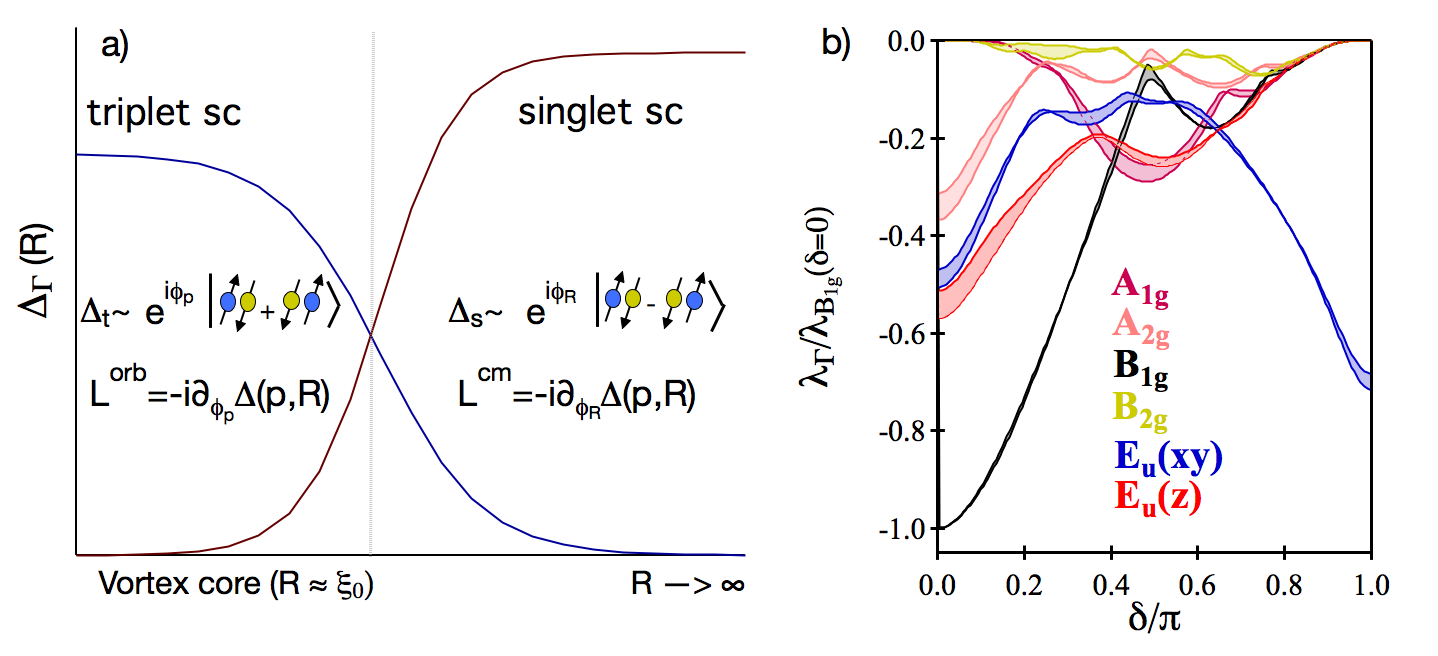}
\caption{[Color online] A schematic picture of the vortex-core state of a d-wave superconductor in the presence of a pairing attraction in a 
triplet channel is shown in panel a). The core state $(R\lesssim \xi_0)$ 
may have a superconducting core with a sizable order-parameter amplitude
with a $p_x\pm i p_y$-symmetry. In panel b) the coupling constants computed using a spin-fluctuation mediated pairing interaction 
equation (\ref{susceptibility}) and a simplified generic band structure of the cuprates. The parameter $\delta$ tunes from predominantly
anti-ferromagnetic ($\delta=0$) to ferromagnetic  ($\delta=\pi$) spin-fluctuations. The width of the curves signifies the span from  
an overdoped ($|\lambda_\Gamma(\delta=0)|$ smaller) to an underdoped ($|\lambda_\Gamma(\delta=0)|$ bigger) material. 
Attraction in the spin-triplet pairing channel is sizable for all $\delta$.  
 }
\label{fig: eigenvalues}
\vspace*{-0.3truecm}
\end{figure*}
 
The scenario is that a finite amplitude of a competing order parameter stabilizes in the vortex-core. An isolated singly-quantized vortex has the asymptotic order parameter $\Delta(\vR)=\Delta_\infty\rm{e}^{i \phi}$ as one circles the core. 
The phase winding of $2\pi$ corresponds to a center-of-mass angular momentum, 
$\hat L^{cm}_z \Delta(\vR)=\frac{\hbar}{i}\frac{\partial}{\partial \phi}\Delta(\vR)= \hbar\Delta(\vR)$,
of the Cooper pairs.  Approaching the vortex core, the order-parameter amplitude is gradually suppressed
and vanishes in the origin of the vortex so to maintain single-valuedness of $\Delta(\vR)$. 
To have a finite order-parameter amplitude in the core 
the center-of-mass angular momentum can rotate in to an internal orbital angular momentum of the Cooper pair, 
$\hat L^{orb}_z\Delta(\vp,\vR)=\frac{\hbar}{i}\frac{\partial}{\partial \phi_{\mathbf{\hat p}}}\Delta(\vp,\vR)= \hbar\Delta(\vp,\vR)$.\cite{sauls2009} 
This scenario occurs in the B-phase of superfluid $^3$He where A-phase and double-core states are found to be energetically favorable to a normal-state core
in different regions of the pressure-temperature phase diagram.\cite{hakonen1989,salomaa1985,thuneberg1987,salomaavolovik1987,fogelstrom1995}    

A d-wave superconductor has singlet-pairing symmetry and to have a finite order parameter in the core with $L^{orb}_z=\hbar$ 
a p-wave order parameter is needed in the vortex-core region. This requires; i) an attractive pairing interaction in a triplet channel, ii) a
symmetry breaking field that introduce a seed of a triplet component. The first condition I argue to be intrinsic in a spin-fluctuation mediated pairing relevant for high-${\rm T_c}$
superconductors (see Ref. \onlinecite{eschrig2006} and references therein). This interaction can support condensation into both spin-singlet and spin-triplet superconducting states. \cite{eschrig2001} 
The second condition is readily given by the weak Zeeman coupling to an external field
always present in a vortex. This is in particular true for extreme type-II superconductors where penetration depth is far larger that the coherence length.
In this paper I neglect orbital effects and the screening of the external magnetic field and assume that the external
magnetic field is constant over the vortex-core region.

\section{Model}
For highly anisotropic spin-fluctuations, 
$\chi^z\equiv\chi_{zz}\gg\chi_{xx,yy}\equiv\chi^\perp$,  
a susceptibility that can be tuned from predominantly antiferromagnetic ($\delta\approx 0$) to
ferromagnetic  ($\delta\approx \pi$) spin-fluctuations can be modeled as
\begin{equation}
\chi^z(\vq) = \sum_{\delta_{x,y}=\pm\delta}
\frac{\chi_Q/4}{1+4\xi_{sfl}^2(\cos^2\frac{q_x-\delta_x}{2}+\cos^2\frac{q_y-\delta_y}{2})}.
\label{susceptibility}
\end{equation} 
$\chi_Q$ is the overall amplitude and
$\xi_{sfl}$ is the spin-spin correlation length which is typically a few lattice constants ($a$) in the cuprates.
Introducing a simple coupling $g$ between 
the spin fluctuations and the quasiparticles the pairing interaction is $V(\vp-\vp^\prime)\equiv V(\vq)=
{\cal{N}}_f g^2 \chi^z(\vq)=\bar \chi^z(\vq)$, ${\cal{N}}_f$ being the total density of states at the Fermi level. The resulting gap equation allows for three channels of pairing, one spin-singlet with $V_s(\vq)=\bar \chi^z(\vq)/2$ and
two spin-triplet channels, one with $V_{tz}(\vq)=\bar \chi^z(\vq)/2$ having the $\vd$-vector parallel
to $\hat z$ and one $V_{t\perp}(\vq)=-\bar \chi^z(\vq)/2$ for which $\vd\perp\hat z$. 
Using equation (\ref{susceptibility}) the gap equation in a weak-coupling approximation reads 
\begin{equation}
\Delta_x(\vpF)=-T\sum_{\vert\epsilon_n\vert \le\epsilon_c}\langle V_x(\vpF-\vpF^\prime)n(\vpF^\prime) 
f_x(\vpF^\prime;\epsilon_n)\rangle_{\vpF^\prime}.
\label{gapequation}
\end{equation}
Here $f_x(\vpF;\epsilon_n), (x=s,tz,t\!\!\perp),$ is the anomalous propagator at Matsubara frequency $\epsilon_n$ and momentum 
$\vpF$ and $n(\vpF)=|\vvF(\vpF)|^{-1}/\langle |\vvF(\vpF^\prime)|^{-1}\rangle_{\vpF^\prime}$.  I use 
a linearized version of eq. (\ref{gapequation}), assuming the factorization $V_x(\vpF-\vpF^\prime)=\sum_\Gamma \lambda_\Gamma {\cal{Y}}_\Gamma(\vpF){\cal{Y}}^*_\Gamma(\vpF^\prime)$, together with 
a tight-binding parameterization of the band structure relevant for BSCCO,\cite{hoogenboom2003} to compute the eigenvalue spectra
for the possible pairing symmetries as a function of $\delta$ keeping $\xi_{sfl}=2 a$.
To each eigenvalue $\lambda_\Gamma$ belongs a set of basis functions
${\cal{Y}}_\Gamma(\vpF)$ which may be classified according to the irreducible representations
$(\Gamma)$ of the crystal group $D_{4h}$. The resulting eigenvalues as function of doping of BSCCO and the degree of incommensuration of the spin fluctuations are shown in figure \ref{fig: eigenvalues}~b. 
Attractive eigenvalues ($\lambda_\Gamma<0$) are found for
the even-parity representations $A_{1g,2g},B_{1g,2g}$ with strongest attraction in the $B_{1g}$-channel ($d_{x^2-y^2}$-wave) followed by the $A_{2g}$-channel (extended s-wave) for dominantly anti-ferromagnetic spin fluctuations. Also
the odd-parity representation $E_{u}$ has attractive eigenvalues. Here it is the channel with $\vd\parallel\hat z$ that is most attractive,  closest in value to that of the $B_{1g}$-channel. This parallel to what was found earlier in the 
case of Sr$_2$RuO$_4$.
\cite{eschrig2001}

\begin{figure*}[t]
\begin{tabular}{lc}	
\hspace{-1.0truecm}\includegraphics[width=1.60\columnwidth,angle=0]{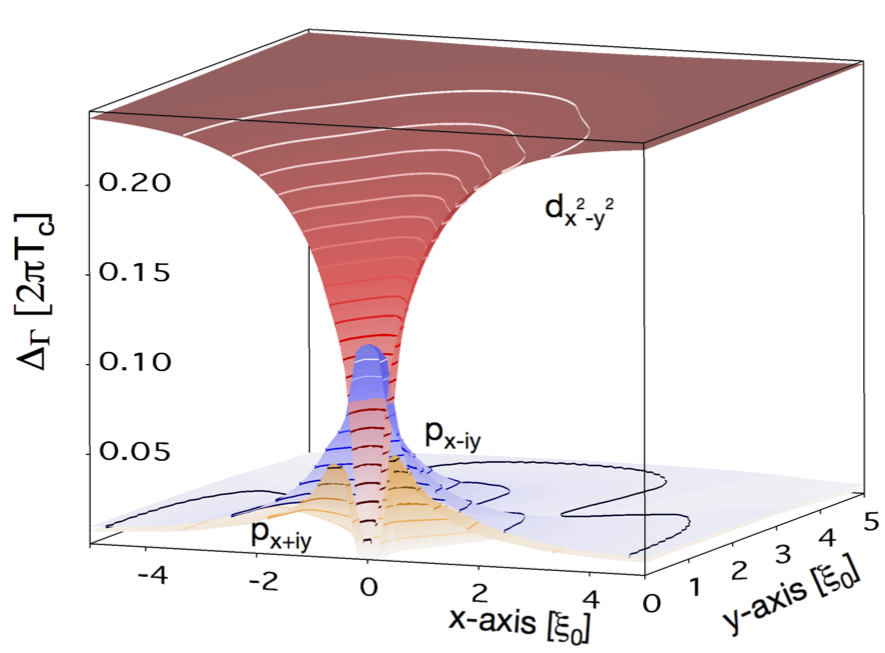}&
\includegraphics[width=0.4\columnwidth,angle=0]{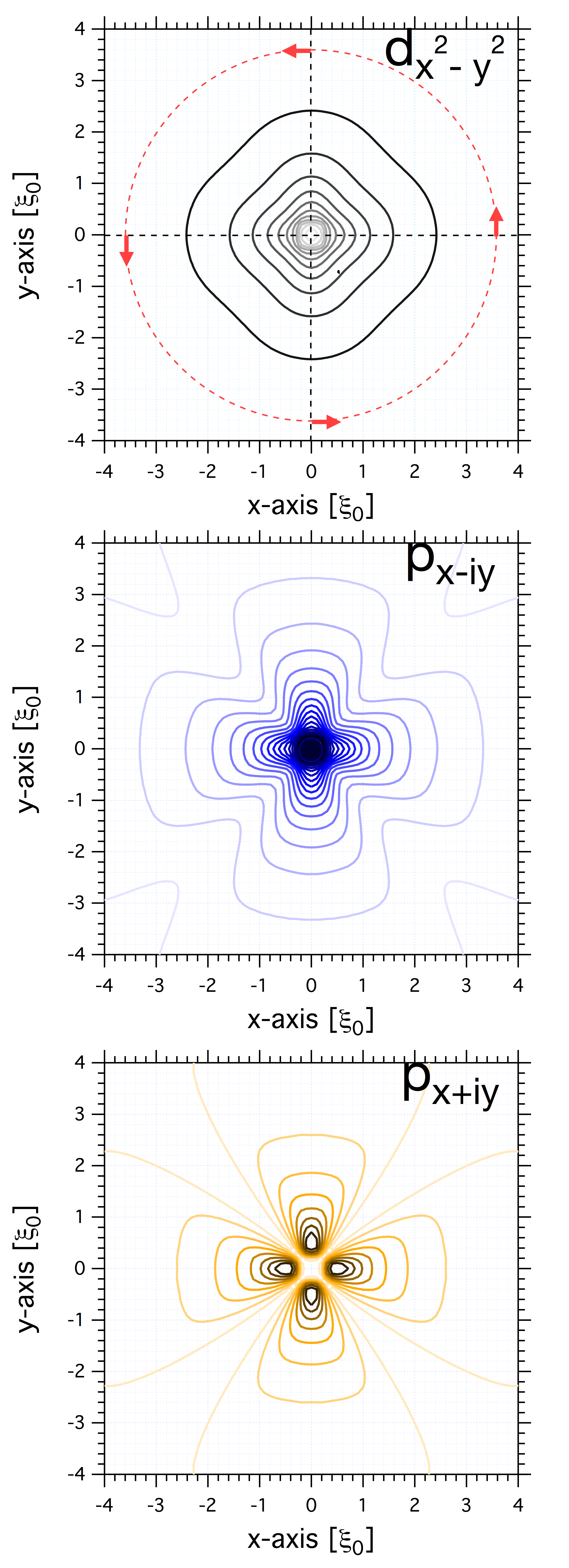}
\end{tabular}
\caption{[Color online] The order-parameter amplitudes computed at $T\!=\!0.05T_{\rm c}$ with 
$\lambda_{E_{u}}\!=\!0.9\lambda_{B_{1g}}$ and $h=0.02\Delta_0$, i.e. 2\% of the zero-temperature
d-wave gap, $\Delta_0$. The asymptotic
d-wave orderparameter having a phase winding of $2\pi$ is as seen suppressed in the core and heals
to its bulk amplitude over roughly 5 $\xi_0$ away from the core.  In the vortex
core a substantial triplet orderparameter $\sim p_{x-iy}$ is stabilized filling an area of
$\sim \pi\xi_0^2$ with an amplitude close to 50\% of the bulk value of the d-wave gap. 
This component has no phase winding but relative phase of $-\pi/2$ to the d-wave component.
Away from the core a second component $\sim p_{x+iy}$ with a phase winding of $4\pi$ appears.  
 }
\label{fig:MixedParityVortex}
\vspace*{-0.2truecm}
\end{figure*}

Next, I need a theory to self-consistently solve for a vortex structure in the presence of a Zeeman field and 
competing order-parameter symmetries and use the quasiclassical theory, a leading order theory in $\Delta/E_F(\ll1)$, 
as presented in e.g. Ref. [\onlinecite{sauls2009}]. 
The central object is the Green's function 
$\hat g(\pfhv,\vR;\varepsilon)=(1/a)\int d\xi_p\hat\tau_3 \hat G (\vp,\vR;\varepsilon),$ obeying the Eilenberger equation
\begin{equation}
i\vvF\!\cdot\!\nabla \hat g(\pfhv,\vR;\varepsilon)
+\lbrack \varepsilon \hat \tau_3\!-\vh\!\cdot \hat\vsigma\!-\hat\Delta(\pfhv,\!\vR),\hat g(\pfhv,\!\vR;\varepsilon)\rbrack\!=\!0,
\label{EE}
\end{equation}
and normalization condition $\hat g(\pfhv,\vR;\varepsilon)^2=-\pi^2$.
The quasiclassical "Hamiltonian", 
$\varepsilon \hat \tau_3\!-\vh\cdot \hat\vsigma\!-\hat\Delta(\pfhv,\vR)$, is a
$4\times4$ matrix in combined particle-hole ($\hat \tau_{i=1,2,3}$) and spin space
($\sigma_{i=x,y,z}$). The order-parameter matrix in (\ref{EE}) reads
\begin{equation}
\hat\Delta(\pfhv,\vR)=\left(\begin{array}{cc} 
0 & \Delta(\pfhv,\vR) \\
\tilde \Delta(\pfhv,\vR)&0\end{array}\right),
\end{equation} 
where $\Delta(\pfhv,\vR)=\lbrack\Delta^{s}(\pfhv,\vR)+\vDelta^{t}(\pfhv,\vR)\!\cdot\!\vsigma \rbrack i\sigma_y$ is a $2\times2$ spin-matrix order parameter.
The superscripts refer to spin-singlet (s) and spin-triplet (t) components of the order parameter. 
Particle-hole components are related via the "tilde"-symmetry $\tilde \alpha(\pfhv,\vR;\varepsilon,t)=\alpha^*(-\pfhv,\vR;-\varepsilon^*,t)$
with $^*$ denoting complex conjugation. This gives
$
\tilde \Delta(\pfhv,\vR)=i\sigma_y \lbrack \Delta^{s*}(-\pfhv,\vR)-\vDelta^{t*}(-\pfhv,\vR)\!\cdot\!\vsigma\rbrack
= i\sigma_y \lbrack\Delta^{s*}(\pfhv,\vR)+\vDelta^{t*}(\pfhv,\vR)\!\cdot\!\vsigma\rbrack.
$
The Zeeman term in (\ref{EE}) reads $\vh\cdot \hat\vsigma={\rm{diag}}\lbrack\vh\cdot \vsigma,-\sigma_y\vh\cdot \vsigma\sigma_y\rbrack$
with the Zeeman field pinning the spin-quantization axis to the z-axis, $\vh\!\cdot\!\vsigma=-\mu_B B \sigma_z$. 
$\vh$ is assumed to be homogeneous and small, $\vert \vh\vert=h \ll \vert \Delta_0\vert$, to weakly break parity
($\Delta_0$ is the zero-temperature d-wave gap).

The Eilenberger equation (\ref{EE}) may be solved by introducing the following two spin-matrix
coherence functions $\gamma=(\gamma_s+\vgamma_t\cdot\vsigma)i\sigma_y$ and $\tilde \gamma=i\sigma_y(\tilde\gamma_s-\tilde\vgamma_t\cdot\vsigma)$
parametrizing the Retarded quasiclassical Green's function, \cite{nagato1993,schopohl1995,sauls2009} 
\begin{equation}
\hat g^{R}=- i \pi \hat N 
\left(\begin{array}{cc}
1+ \gamma \tilde \gamma & 2 \gamma \\
-2 \tilde \gamma &-1 -\tilde \gamma \gamma \end{array}\right)
=\left(\begin{array}{cc}g&f\\ 
\tilde f&\tilde g\end{array}\right),
\label{qcgmatrix}
\end{equation}
with $\hat N={\rm diag}[(1- \gamma \tilde \gamma)^{-1},(1- \tilde \gamma \gamma)^{-1}]$.
The Advanced function is given as $\hat g^A=\tau_3 \hat g^{R\dagger}\tau_3$ and the Matsubara function as 
$\hat g^M(\epsilon_n)=\hat g^R(\epsilon+i0\rightarrow i\epsilon_n)$.
The mixed-parity orderparameter components are linear combinations of the
singlet and the z-component of the triplet part as $\Delta_\pm(\pfhv,\vR)=\pm\lbrack\Delta^s(\pfhv,\vR)\pm\Delta^t_z(\pfhv,\vR)\rbrack$
(and $\tilde\Delta_\pm(\pfhv,\vR)=\pm\lbrack\Delta^{s*}(\pfhv,\vR)\pm\Delta^{t*}_z(\pfhv,\vR)\rbrack$).
This leads, together with the Zeeman-shifted frequency $\varepsilon_\pm=\varepsilon\pm\mu_B B$, to a separation in to two
pseudo-spin bands ($\pm$) with different orderparameters $\Delta_\pm(\pfhv,\vR)$ and equation (\ref{EE}) is written as two pairs of scalar Riccati equations
\begin{eqnarray} 
i\vvF\!\cdot\!\nabla \gamma_\pm +2 \varepsilon_\pm \gamma_\pm&=&-\gamma_\pm \tDelta_\pm \gamma_\pm -\Delta_\pm
\label{gamman}\\
i\vvF\!\cdot\!\nabla \tilde \gamma_\pm -2 \varepsilon_\pm \tilde \gamma_\pm&=&-\tilde \gamma_\pm \Delta_\pm \tilde \gamma_\pm-\tDelta_\pm
\label{gammat}
\end{eqnarray}
one for each spin-band. The two equations are solved by numerical integration
along straight lines, or trajectories, $\vs(x)=\vs_0\pm x\, \vvF/\vert\vvF\vert$ for $\gamma_\pm$ and  $\tilde \gamma_\pm$ as described in e.g. Ref. [\onlinecite{sauls2009}]. Once $\gamma_\pm(\pfhv,\vR;\epsilon)$Êand $\tilde \gamma_\pm(\pfhv,\vR;\epsilon)$ are obtained, the  
order-parameter fields are calculated using (\ref{gapequation}) as
\begin{eqnarray}
\Delta^s_d(\vR)\!&=&\!-\lambda_{B_{1g}} T\sum_{|\epsilon_n|\le\epsilon_{c}} \langle {\cal{Y}}_{B_{1g}}^*(\pfhv)\frac{(f_+-f_-)}{2}
\rangle_{\pfhv}
\label{OPs}\\
\Delta^t_{p_x\pm ip_y}(\vR)\!&=&\!-\lambda_{E_{u}} T\sum_{|\epsilon_n|\le\epsilon_{c}} \langle {\cal{Y}}_{E_{u},\pm}^*(\pfhv)\frac{(f_++f_-)}{2}
\rangle_{\pfhv}
\label{OPt}
\end{eqnarray}
where $f_\pm=f_\pm(\pfhv,\vR;\epsilon_n)$ are the anomalous functions on spin-band $\pm$. $\langle \cdots \rangle_{\pfhv}=\int\frac{d\phi_p}{2\pi}$ is the
average over the momentum direction $\pfhv$ on the Fermi surface, 
with the angle
$\phi_p$ giving the angle the momentum $\pfhv$ makes to the crystal a-axis (x-axis in the figures). In the vortex calculations the conventional basis functions 
${\cal{Y}}_{B_{1g}}(\pfhv)=\sqrt{2}\cos2\phi_p$ and ${\cal{Y}}_{E_{u},\pm}(\pfhv)=\sqrt{2}(\cos\phi_p\pm i \sin\phi_p)$ are used. The paring interaction and cut-off frequency, $\epsilon_c$,
are eliminated in favor of the transition temperature $T_c$ as $-\lambda_{B_{1g}}^{-1}=\ln{T/T_c}+\sum_{n\ge0}^{n\le c}(n+1/2)^{-1}$. The subdominant interaction $\lambda_{E_{u}}$ is introduced in the self-consistent calculations as fraction of the dominant 
one and is treated as a parameter free to explore. 

\begin{figure}[t]
\includegraphics[width=1\columnwidth,angle=0]{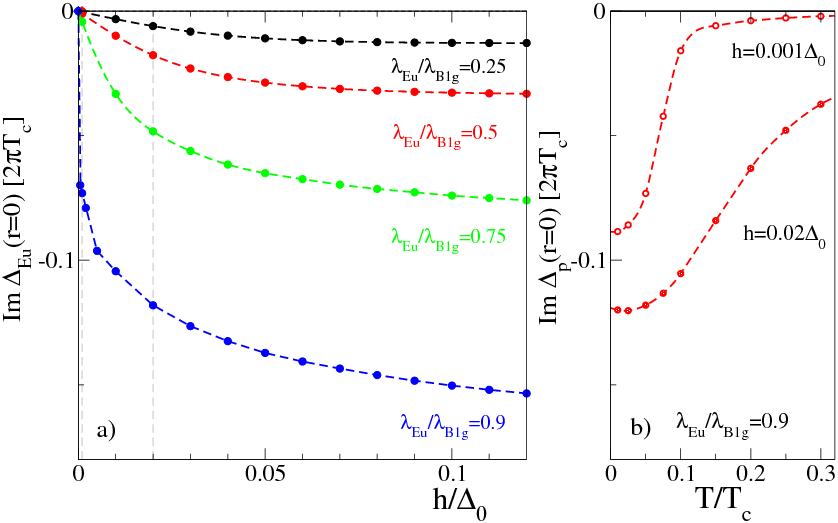}
\caption{[Color online] The amplitude of the p-wave order parameter in the vortex core, $\Delta_{p}(0)$, computed at $T=0.05T_c$, is shown in panel a)
as a function of the symmetry-breaking field $h$ for various
ratios $\lambda_{E_{u}}/\lambda_{B_{1g}}$. 
For large ratios, $\lambda_{E_{u}}/\lambda_{B_{1g}} \lesssim 0.9$,
the p-wave stabilizes even as $h\rightarrow0$ (but $h$ finite). The temperature dependence of $\Delta_{p}(0)$
is shown in panel b) at $h=0.001 \Delta_0$ and at $h=0.02 \Delta_0$. For the larger field $\Delta_{p}(0;T)$ is finite at
higher temperatures and grows with decreasing temperature while for $h=0.001 \Delta_0$, $\Delta_{p}(0;T)$
has a distinct temperature below which it grows rapidly to its low-T value.
For smaller ratios, $\lambda_{E_{u}}/\lambda_{B_{1g}}\lesssim 0.7$,
the p-wave core phase is close to linearly dependent on $h$.
}
\label{fig:op_vs_h}
\vspace*{-0.2truecm}
\end{figure}

\section{Results}
Equations (\ref{gamman},\ref{gammat}) and (\ref{OPs},\ref{OPt}) are iterated until self-consistency is reached.  
In figure \ref{fig:MixedParityVortex}, the structure of a mixed-parity vortex is shown. A
substantial triplet p-wave order parameter may be nucleated in the singlet d-wave vortex core 
with both p-wave components $p_x\pm i p_y$ present.
The p-wave component, $\Delta_p$, with $L^{orb}_{z,p-wave}=L^{cm}_{z,d-wave}$ will be finite in the core center and it carries no phase winding. The p-wave component, $\Delta_{p^\prime}$, with $L^{orb}_{z,p^\prime-wave}=-L^{cm}_{z,d-wave}$ has a finite amplitude on the phase boundary
separating the singlet and triplet order parameters, at 
$R\sim \xi_0=\hbar v_F/2\pi T_{\rm c}$, around which its phase winds by $4\pi$. This so that 
$L^{orb}_{z,p^\prime-wave}+L^{cm}_{z,p^\prime-wave}=L^{cm}_{d-wave}$. 
All amplitudes retain the four-fold symmetry of the $d_{x^2-y^2}$ amplitude as seen in the contour plots displayed in figure \ref{fig:MixedParityVortex}.

The nucleation of a p-wave order parameter is dependent on a finite Zeeman field. In figure \ref{fig:op_vs_h} the amplitude
$\Delta_p(0)$ is displayed as a function of $h$ for different strength of $\lambda_{E_{u}}$. $\Delta_p(0)$ is finite for all coupling strengths
and grows with increasing Zeeman field. For larger couplings,  $\lambda_{E_{u}}\gtrsim 0.7 \lambda_{B_{1g}}$, the onset of
$\Delta_p(0)$ at small fields becomes increasingly nonlinear with sharp onset of the sub-dominant order parameter at $h\gtrsim0$. 
In panel b) of  figure \ref{fig:op_vs_h} the temperature dependence of
$\Delta_p(0)$ is shown. While the p-wave amplitude is finite at all temperatures there is a transition from a field induced
triplet order parameter at larger fields, when $h\gtrsim 0.02 \Delta_0$, and at high temperatures $T\gtrsim 0.1T_c$, to an intrinsic
phase transition at vanishingly small fields $h\lesssim 0.01 \Delta_0$ and low temperatures. For weaker triplet pairing strength
the intrinsic phase transition in to a $d+ip$ vortex state is pushed to lower temperatures.

\begin{figure}
\hspace{-0.5truecm}\includegraphics[width=.99\columnwidth,angle=0]{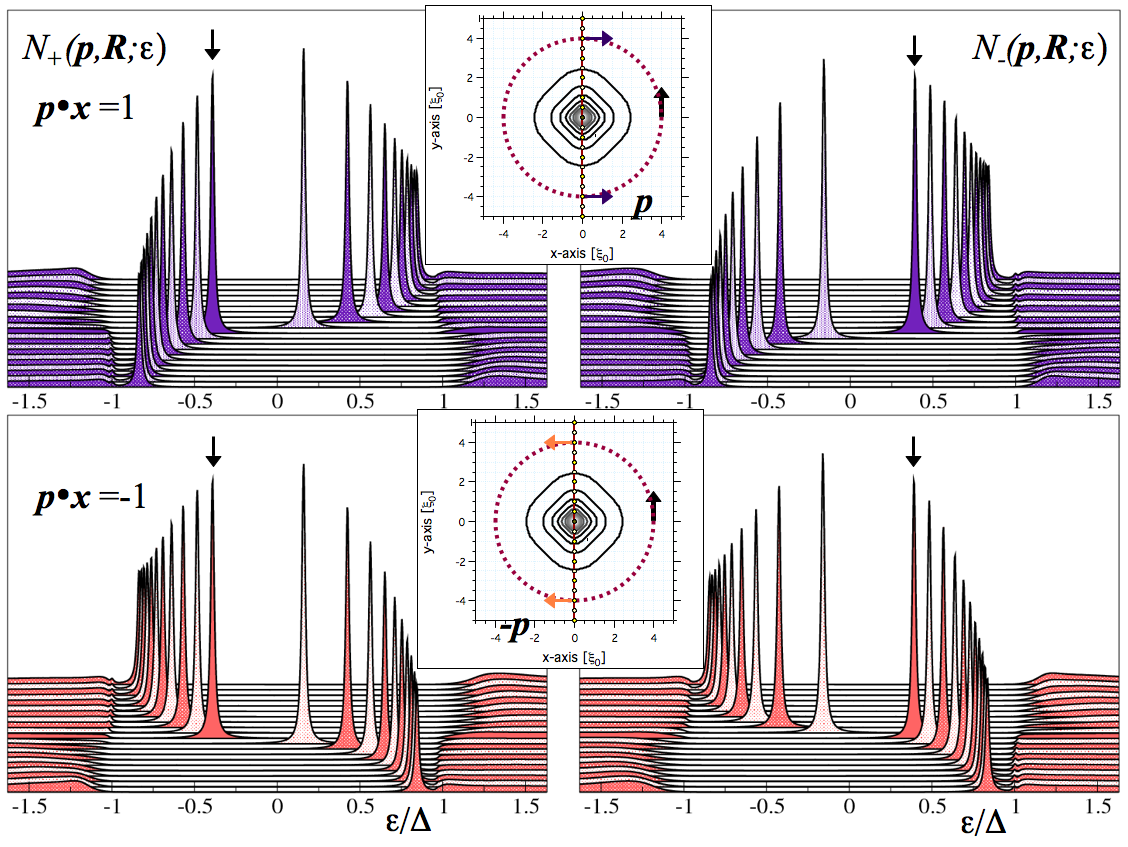}
\caption{[Color online] The trajectory-resolved DoS $N_{\rm \pm} (\pfhv,\vR;\epsilon)$ evaluated on the cut along the y-axis through the origin. The spectra are taken in the indicated points separated by $0.5\xi_0$ (see the inserts). The phase winding of the d-wave is counter clock wise in this case. The direction of $\pfhv$ is chosen (anti-)parallel to the x-axis in (lower) upper pair of panels. The presence of a subdominant order-parameter amplitude introduces a shift of the the state in the core center to a finite energy (this state is marked by an arrow in each panel). This energy shift is negative (positive) for the branch $N_+ (N_-)$. There still exist zero-energy states on both branches, but these are now found away from the core center. 
}
\label{fig:TrajDoS}
\vspace*{-0.2truecm}
\end{figure}

\begin{figure*}
\hspace{-0.5truecm}\includegraphics[width=2\columnwidth,angle=0]{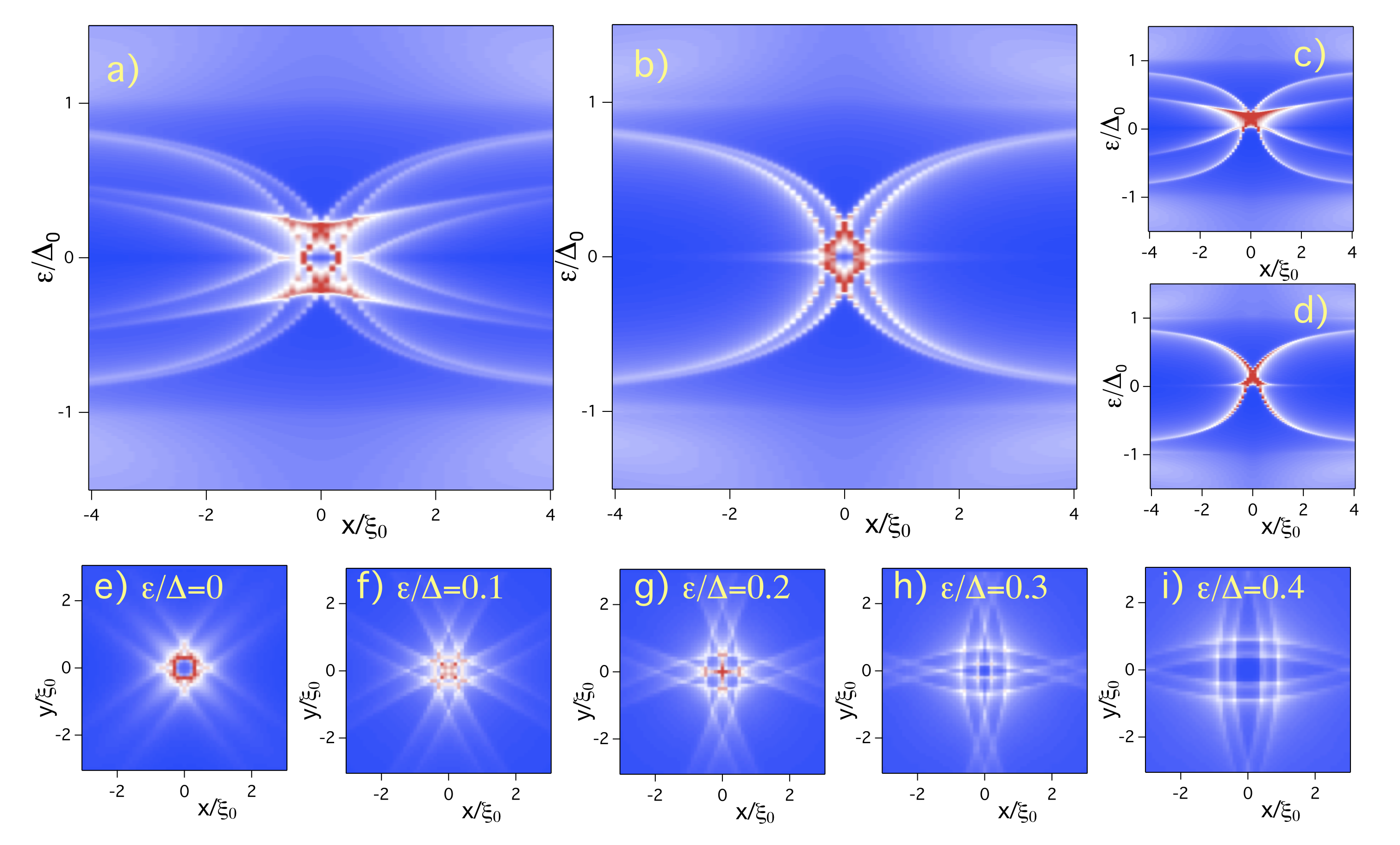}
\caption{[Color online] The DoS of a vortex with a finite p-wave core phase. In panels a) and b) $N_{\rm Tot}(\vR;\epsilon)$ is displayed
as function of distance from the core in a) along the $y-axis$ at $x=0$ and in b) along $x=y$.  
In panels (c-d) as in (a-b) the DoS $N_{\rm -}(\vR;\epsilon)$ for one spin-band.
There is a dispersion of the core states as a function of distance from the core but as seen in the spin-resolved DoS,
the total DoS also consists of two non-dispersing states with large spectral weight located at $\epsilon=\pm 0.2 \Delta_0$. 
In panels (e-i) a spatial map of the DoS is shown at various $\epsilon$. 
 }
\label{fig:DoS}
\vspace*{-0.2truecm}
\end{figure*}

The emergence of a mixed-parity state in the vortex center at low temperatures has a profound effect on the quasiparticle spectra. 
In general the spatially resolved density of states (DoS), 
$N_{\rm Tot} (\vR;\epsilon)=-\frac{1}{\pi}{\rm Im} \langle {\rm Tr} [ \hat\tau_3\hat g^R(\pfhv,\vR;\epsilon+i0^+) ]\rangle_{\pfhv}$, will show
evidence of the sub-gap Caroli-deGennes-Matricon states that carry the screening current of a vortex. \cite{caroli1964,rainer1996} 
Resolving the spectra also in position on the Fermi surface $\pfhv$ one finds for a pure d-wave vortex, 
on trajectories tangential to the asymptotic phase winding,  
that the bound states in the core have the {\em qualitative} dispersion with distance or `impact parameter' $b$ from the vortex-core  
$E(\pfhv,b)\approx\pm |\Delta(\pfhv)|\tanh (b/b_0)$. Here, $b_0\approx \xi_0$ is a scale factor
and $\pm$ denotes if the momentum direction is parallel (+) or anti-parallel (-) to the phase winding of the asymptotic order parameter 
$\Delta(\pfhv) {\rm e}^{i\phi}$ at $b\ll0$.  Introducing an imaginary p-wave order parameter in the core splits the quasiparticle spectra in to two branches, one for each 
spin band $\alpha(=\pm)$. This is displayed in figure \ref{fig:TrajDoS}. The qualitative quasiparticle dispersion of the core states is now modified as $E_\alpha(\pfhv,b)\approx\pm |\Delta(\pfhv)| \tanh [(b\mp s_\alpha\bar b)/b_0]$
where $s_\pm=\pm 1$. The offset, $\bar b$, is a direct
consequence of a finite order parameter in the vortex core center and
shifts the zero-energy state to a finite impact parameter away from the core center. In the core center 
the quasiparticle state is shifted to finite energy, $E_\alpha(\pfhv,0)\approx -s_\alpha |\Delta(\pfhv)| \tanh (\bar b/b_0)$. 
Note that both states, with and against the phase winding $(\pm)$ on one branch are shifted to the same energy. This
leads to a suppression of the screening current in the core area,  $|\vR|\lesssim 1\xi_0$.

The features of the trajectory resolved DoS are detectable in the trajectory averaged total DoS, $N_{\rm Tot} (\vR;\epsilon)$, which is directly related to the tunneling conductance measured by STS.\cite{fischer2007} In figure
\ref{fig:DoS} the DoS calculated at $T=0.025T_c$ with $\lambda_{E_{u}}=0.9\lambda_{B_{1g}}$ and $h=0.02\Delta_0$
is shown. $N_{\rm Tot} (\vR;\epsilon)$ on a ray through the vortex core along an anti-node (panel a)  
lacks a zero energy state (ZES) in the core center. The core state is pushed to $E_\alpha(0)\approx\mp 0.3\Delta_0$ and this state have very little dispersion with position on the ray. ZES are found at a distance $b\approx \pm 0.5\xi_0$ from the core center
with half the spectral weight of the ZES in a pure d-wave vortex as the two spin bands are shifted differently by $\Delta_p(0)$.
The features in the DoS are also generally broadened by the angle average $\langle\cdots \rangle_{\pfhv}$.
The corresponding spin-band resolved DoS is shown in panel c for spin band (-).  
On a ray through the vortex core along a node (panel b) the DoS also lack ZES in the core center but the core state
$E_\alpha(0)\approx\mp 0.3\Delta_0$ has more of a dispersion with small $b$ compared to that in the anti-nodal direction. 
This is due to the linear opening of the energy gap around the node,
 $|\Delta_d(\phi)|\sim |(\partial \Delta/\partial \phi) (\phi-\phi_{node})|$ of the d-wave gap. In panels (e-i) in figure \ref{fig:DoS} the spatially resolved
DoS is displayed on a  $6\xi_0\times 6\xi_0$-square with center in the vortex core at different fixed energies. The ZES form ring around the core center and the cores states, $E_\alpha(0)\approx\mp 0.3\Delta_0$, extends along the anti-nodes. The overall shape of the vortex is a doubling of the star-shaped
DoS found in the pure d-wave vortex core. \cite{schopohl1995} The doubling shows up as a square lattice in the DoS amplitude, and the
lattice constant is set by the magnitude of the induced triplet order parameter, $\Delta_p(0)$.  

\section{Concluions}
In closing, I have shown that a mixed-parity $d+ip$-vortex state is possible to stabilize in a high-T$_c$ superconductor. The weak Zeeman coupling to the external magnetic field gives a sufficient seed to nucleate a p-wave order parameter in the vortex core. The necessary attractive triplet-pairing channel is supported by spin-fluctuation mediated pairing, argued relevant for the cuprates.
This new core state is directly detectable in STS measurements of the quasiparticle spectra and I find, within
the limits of quasiclassical theory, good agreement with existing experimental data.\cite{maggioaprile1995,pan2000,hoffman2002,levy2005,fischer2007}
A further study using the Bogoliubov-de Gennes equations is needed to quantitatively compare the theory with experiments. This as the quasiclassical theory does not self-consistently resolve the angular-momentum quantization of the Caroli-deGennes-Matricon states. This quantization gives a finite shift, or a mini gap, of the lowest energy state from the Fermi surface  $\sim\Delta^2/E_F$.\cite{caroli1964,gygi1991} For the 
high-$T_{\rm c}$ cuprates this mini gap may be sizable as $\Delta/E_F\approx 0.1$ and may very well remove the states found at $E=0.0$ away from the core center.       
    
{\em Acknowledgment:} Interaction with {\O}ystein Fischer during this work, from the initial curious questions
to discussing results, has been extremely valuable. 

\end{document}